\documentclass[aps,prd,twocolumn,nofootinbib,showpacs]{revtex4-1}

\usepackage{graphicx}

\usepackage[colorlinks=true, linkcolor=blue, citecolor=blue, urlcolor=blue]{hyperref}

\begin{document}

\title{Novel All-Orders Single-Scale Approach to QCD Renormalization Scale-Setting}

\author{Jian-Ming Shen$^1$}
\email{cqusjm@cqu.edu.cn}
\author{Xing-Gang Wu$^1$}
\email{wuxg@cqu.edu.cn}
\author{Bo-Lun Du$^1$}
\email{dblcqu@cqu.edu.cn}
\author{Stanley J. Brodsky$^2$}
\email{sjbth@slac.stanford.edu}

\affiliation{$^1$Department of Physics, Chongqing University, Chongqing 401331, P.R. China}
\affiliation{$^2$SLAC National Accelerator Laboratory, Stanford University, Stanford, California 94039, USA}

\begin{abstract}

The Principle of Maximal Conformality (PMC) provides a rigorous method for eliminating renormalization scheme-and-scale ambiguities for perturbative QCD predictions. The PMC uses the renormalization group equation to fix the $\beta$-pattern of each order in an arbitrary pQCD approximant, and it then determines the optimal renormalization scale by absorbing all $\{\beta_{i}\}$ terms into the running coupling at each order. The resulting coefficients of the pQCD series match the scheme-independent conformal series with $\beta=0$. As in QED, different renormalization scales appear at each order; we call this the multi-scale approach. In this paper, we present a novel single-scale approach for the PMC, in which a single effective scale is constructed to eliminate all non-conformal $\beta$-terms up to a given order simultaneously. The PMC single-scale approach inherits the main features of the multi-scale approach; for example, its predictions are scheme-independent, and the pQCD convergence is greatly improved due to the elimination of divergent renormalon terms. As an application of the single-scale approach, we investigate the $e^+e^-$ annihilation cross-section ratio $R_{e^+e^-}$ and the Higgs decay-width $\Gamma(H \to b \bar{b})$, including four-loop QCD contributions. The resulting predictions are nearly identical to the multi-scale predictions for both the total and differential contributions. Thus in many cases, the PMC single-scale approach PMC-s, which requires a simpler analysis, could be adopted as a reliable substitution for the PMC multi-scale approach for setting the renormalization scale for high-energy processes, particularly when one does not need detailed information at each order. The elimination of the renormalization scale uncertainty increases the precision of tests of the Standard Model at the LHC.

\end{abstract}


\maketitle

\section{Introduction}

A primary requirement of renormalization group invariance (RGI) is that a valid prediction for a physical observable from quantum field theory must be independent of the choice of renormalization scheme, such as the minimum-subtraction $\overline{\rm MS}$ or ${\rm  MOM}$ schemes. Conventional predictions based on a truncated perturbation series do not automatically satisfy this requirement, leading to scheme-and-scale ambiguities. For example, perturbative QCD predictions, where the renormalization scale and its range are simply guessed, lead to an unphysical dependence on the choice of renormalization scheme.

The renormalization group equation (RGE) provides a rigorous basis for determining the running behavior of the coupling constant and hence the setting of the renormalization scale. It determines the running of the strong coupling from the analytic properties of the $\beta$-function:
\begin{equation}
\beta(\alpha(\mu))=\mu^2\frac{{\rm d} \alpha(\mu)} {{\rm d}\mu^2} = -\alpha^2(\mu) \sum_{i=0}^\infty \beta_i \alpha^{i}(\mu), \label{rge}
\end{equation}
where a  perturbative expansion of the $\beta$-function in terms of $\alpha=\alpha_s/4\pi$ is assumed. The ``Principle of Maximal Conformality" (PMC)~\cite{Brodsky:2011ta, Brodsky:2011ig, Mojaza:2012mf, Brodsky:2013vpa} utilizes the RGE recursively to unambigously identify the occurrence and pattern of nonconformal $\{\beta_i\}$ terms at each order in a pQCD expansion. The PMC then determines the optimal renormalization scales  by absorbing all occurrences of the $\{\beta_i\}$ terms into the scales of the running coupling at each order of perturbation theory. The coefficients of the resulting pQCD series then match the ``conformal" series with $\beta=0$. Given one measurement which sets the value of the coupling at a scale, the resulting PMC predictions are independent of the choice of renormalization scheme. Thus the PMC scale setting eliminates an unnecessary theoretical uncertainty. There is another uncertainty from different choices of factorization scale, whose determination is a separate issue, which may be solved by matching nonpertubative bound-state dynamics with perturbative DGLAP evolution~\cite{Gribov:1972ri, Altarelli:1977zs, Dokshitzer:1977sg} \footnote{We have observed that the factorization scale dependence could be suppressed after applying the PMC~\cite{Wang:2014sua, Wang:2016wgw}, which may be explained by the fact that the pQCD series behaves much better after applying the PMC.}.

The elimination of the renormalization scale uncertainty for pQCD is important since it increases the precision of tests of the Standard Model at the LHC. The scales predicted by the PMC are physical -- they reflect the virtualities of the gluon propagators at each given order, as well as setting the effective number of active flavors $n_f$. Specific renormalization scales and values of $n_f$ appear for each skeleton graph.  The QCD scales determined by the PMC can thus be considered as the relevant physical scales for observables, in analogy to QED. In fact, the PMC method reduces in the Abelian limit to the standard Gell Mann-Low method for setting the renormalization scale for precision predictions in QED~\cite{GellMann:1954fq}.

In practice, the PMC multi-scale method requires considerable theoretical analysis. In this paper, we introduce a new all-orders single-scale approach ``PMC-s"  which makes the implementation and automation of PMC scale-setting simpler and more transparent. In effect, the PMC-s provides a mean value for the PMC multi-scales, while retaining its central predictions. We also find that the single PMC-s scale shows stability and convergence with increasing order in pQCD.

The remaining parts of this paper are organized as follows. We will give the PMC single-scale method in Sec.II. We will then apply it to two examples, i.e. the $R$-ratio at the $e^+ e^-$ collider and the decay of $H\to b\bar{b}$, in Sec. III. Section IV is reserved for a summary.

\section{Calculation technology}

As we have shown in our previous papers~\cite{Brodsky:2013vpa, Mojaza:2012mf, Bi:2015wea}, the $\{\beta_i\}$-dependence of any pQCD expression occurs with a specific ``degeneracy" pattern dictated by the RGE. Specifically, one finds
\begin{widetext}
\begin{eqnarray}
\rho(Q) &=& r_{1,0}{\alpha(\mu)^p} + \left[ r_{2,0} + p \beta_0 r_{2,1} \right]{\alpha(\mu)^{p+1}} + \left[ r_{3,0} + p \beta_1 r_{2,1} + (p+1){\beta _0}r_{3,1} + \frac{p(p+1)}{2} \beta_0^2 r_{3,2} \right]{\alpha(\mu)^{p+2}} \nonumber\\
&& + \bigg[ r_{4,0} + p{\beta_2}{r_{2,1}} + (p+1){\beta_1}{r_{3,1}} + \frac{p(3+2p)}{2}{\beta_1}{\beta_0}{r_{3,2}} + (p+2){\beta_0}{r_{4,1}} + \frac{(p+1)(p+2)}{2}\beta_0^2{r_{4,2}} \nonumber\\
&& + \frac{p(p+1)(p+2)}{3!}\beta_0^3{r_{4,3}} \bigg]{\alpha(\mu)^{p+3}} + \cdots,
\end{eqnarray}
\end{widetext}
where $r_{1,0}$ is the tree-level term and $p$ is the power of the coupling associated with the tree-level term, $\mu$ is the initial renormalization scale, and $Q$ represents the kinematic scale. The pattern of $\{\beta_i\}$ terms from one order to the next are general properties of non-Abelian gauge theory for any physical observable.

The pQCD expansion for $\rho(Q)$ can be reorganized into the following compact form:
\begin{widetext}
\begin{eqnarray}
\rho(Q) = \sum\limits_{n \ge 1} {r_{n,0}}{\alpha(\mu)^{n+p-1}} + \sum\limits_{n \ge 1} \left[(n+p-1)\alpha(\mu)^{n+p-2}\beta\right] \sum\limits_{j \ge 1}(-1)^{j} \Delta_{n}^{(j-1)} r_{n+j,j}
\label{eq:generalization}
\end{eqnarray}
\end{widetext}
where we have introduced the notation:
\begin{eqnarray}
\Delta_n^{(0)} &=& 1, \nonumber\\
\Delta_n^{(1)} &=& \frac{1}{2!}\left[\frac{\partial\beta}{\partial\alpha}+ (n+p-2)\frac{\beta}{\alpha}\right], \nonumber\\
\Delta_n^{(2)} &=& \frac{1}{3!}\left[\beta \frac{\partial^2 \beta}{(\partial\alpha)^2} + \left(\frac{\partial\beta}{\partial\alpha}\right)^2 + 3(n+p-2)\frac{\beta}{\alpha}\frac{\partial\beta}{\partial\alpha}\right. \nonumber\\
&& \left. \quad\quad+ (n+p-2)(n+p-3)\frac{\beta^2}{\alpha^2} \right], \nonumber\\
&& \cdots  \nonumber
\end{eqnarray}
As a further step, we can explicitly identify the scale dependence of the non-conformal coefficients $r_{i,j(\geq1)}$ as
\begin{eqnarray}
r_{i,j} &=& \sum_{k=0}^{j} C_j^k L^k \hat{r}_{i-k,j-k},
\label{eq:rijLogPart}
\end{eqnarray}
where $L=\ln(\mu^2/Q^2)$, $\hat{r}_{i,j}=r_{i,j}|_{\mu=Q}$, and the combinatorial coefficients are $C_j^k={j!}/{k!(j-k)!}$. The conformal coefficients are free from scale dependence; i.e., $r_{i,0}=\hat{r}_{i,0}$. By substituting Eq.(\ref{eq:rijLogPart}) into Eq.(\ref{eq:generalization}), we obtain
\begin{widetext}
\begin{eqnarray}
\rho(Q) =&& \sum\limits_{n \ge 1} {\hat{r}_{n,0}}{\alpha(\mu)^{n+p-1}} + \sum\limits_{n \ge 1} \left[(n+p-1)\alpha(\mu)^{n+p-2}\beta\right] \sum\limits_{j \ge 1}(-1)^{j} \Delta_{n}^{(j-1)} \hat{r}_{n+j,j} \nonumber\\
&& + \sum\limits_{k \ge 1} L^k\sum\limits_{n \ge 1} \left[(n+p-1)\alpha(\mu)^{n+p-2}\beta\right]\sum\limits_{j \ge k}(-1)^{j} C_{j}^{k} \Delta_{n}^{(j-1)} \hat{r}_{n+j-k,j-k}.
\label{eq:rhomu}
\end{eqnarray}
\end{widetext}

Following the PMC procedure, all non-conformal terms should be resummed into the running coupling. In the case of the multi-scale approach, one can do this recursively, leading to a scale-fixed scheme-independent conformal series~\cite{Brodsky:2013vpa, Mojaza:2012mf}:
\begin{equation}
\rho(Q)=\sum\limits_{n \ge 1} {\hat{r}_{n,0}}{\alpha(Q_n)^{n+p-1}},
\end{equation}
where $Q_n$ are the PMC scales appearing at each orders. The PMC scales depend on the choice of renormalization scheme; however, once the value of $\alpha_s (Q)$ is determined in the chosen scheme at a specific physical kinematic scale $Q$, the resulting PMC predictions are independent of the scheme choice.

In the following, we shall show that by introducing a single universal renormalization scale $Q_\star$, one can also obtain a scheme-independent conformal series, i.e.
\begin{eqnarray}
\rho(Q) = \sum\limits_{n \ge 1} {\hat{r}_{n,0}}{\alpha(Q_\star)^{n+p-1}}.
\end{eqnarray}
This can be achieved by replacing the scale $\mu$ in Eq.(\ref{eq:rhomu}) as $Q_\star$, whose value is determined by requiring all non-conformal terms vanish. The solution of $\ln{Q_{\star}^2}/{Q^2}$ can be written as a power series in $\alpha(Q)$, i.e.
\begin{equation}
\ln\frac{Q_{\star}^2}{Q^2} = T_0 + T_1 \alpha(Q) + T_2 \alpha^2(Q)+\cdots,  \label{eq:PMCscale}
\end{equation}
where $T_i$ are process-dependent coefficients. The coefficients $T_i$ $(i=0,1,\cdots,n)$ can be fixed by a N$^{n+1}$LO pQCD calculation. For example, for a N$^3$LO calculation, we can get a next-to-next-to-leading log order (N$^2$LL) $Q^*$, whose three coefficients are
\begin{widetext}
\begin{eqnarray}
T_0 &=& -\frac{\hat{r}_{2,1}}{\hat{r}_{1,0}}, \label{eq:t0}\\
T_1 &=& \frac{(p+1)(\hat{r}_{2,0} \hat{r}_{2,1}- \hat{r}_{1,0} \hat{r}_{3,1})}{p \hat{r}_{1,0}^2}+\frac{(p+1)(\hat{r}_{2,1}^2-\hat{r}_{1,0} \hat{r}_{3,2})}{2 \hat{r}_{1,0}^2}\beta_0, \label{eq:t1}\\
T_2 &=& \frac{(p+1)^2 \left(\hat{r}_{1,0} \hat{r}_{2,0} \hat{r}_{3,1} - \hat{r}_{2,0}^2 \hat{r}_{2,1} \right)
+ p(p+2) \left(\hat{r}_{1,0} \hat{r}_{2,1} \hat{r}_{3,0}-\hat{r}_{1,0}^2 \hat{r}_{4,1}\right)}{p^2 \hat{r}_{1,0}^3}+\frac{(p+2)\left(\hat{r}_{2,1}^2-\hat{r}_{1,0}\hat{r}_{3,2}\right)}{2\hat{r}_{1,0}^2}\beta_1  \nonumber\\
&&-\frac{p(p+1)\hat{r}_{2,0} \hat{r}_{2,1}^2+(p+1)^2 \left(\hat{r}_{2,0} \hat{r}_{2,1}^2- 2\hat{r}_{1,0}\hat{r}_{2,1}\hat{r}_{3,1}-\hat{r}_{1,0} \hat{r}_{2,0} \hat{r}_{3,2}\right)+(p+1)(p+2)\hat{r}^2_{1,0} \hat{r}_{4,2}}{2p\hat{r}_{1,0}^3}\beta_0 \nonumber\\
&&+\frac{(p+1)(p+2)\left(\hat{r}_{1,0}\hat{r}_{2,1}\hat{r}_{3,2}-\hat{r}_{1,0}^2 \hat{r}_{4,3}\right)+(p+1)(1+2p)\left(\hat{r}_{1,0} \hat{r}_{2,1} \hat{r}_{3,2}-\hat{r}_{2,1}^3\right)}
{6\hat{r}_{1,0}^3}\beta_0^2.  \label{eq:t2}
\end{eqnarray}
\end{widetext}
It is interesting that different orders of the perturbative series for the PMC scale have an identical form; e.g., the coefficients of $(p+i+1)\beta_i\alpha^{i+1}(Q)$ are the same. Moreover, the effective scale $Q_\star$ is explicitly independent of the choice of initial choice of the renormalization scale $\mu$ at any fixed order. It thus has universal properties. It also converges rapidly as shall be shown below; thus any residual scale dependence due to uncalculated higher-order terms is greatly suppressed. Another important feature is that the single-scale approach avoids the problem of very small arguments of the running coupling appearing at a specific order; e.g., when a soft gluon carries the momentum flow. An example of this appears in the analysis of the Bjorken sum rule~\cite{BjorkenPMC}. On the other hand, in some leading-twist processes such as single spin asymmetries in deep inelastic scattering~\cite{Brodsky:2002cx} or the double Boer-Mulders effect in lepton pair production~\cite{Boer:2002ju}, the scale of the running coupling at specific orders will be soft since these processes involve gluonic initial-state or final-state interactions at small momentum transfer.

A related single-scale approach has been suggested in Refs.\cite{Grunberg:1991ac, Brodsky:1995tb} by applying the Brodsky-Lepage-Mackenzie scale-setting approach~\cite{Brodsky:1982gc} \footnote{In those two references, only two-loop expressions are given, but we have found that such an approach can be extended to all orders. A detailed discussion on this point is in preparation.}. However, in these analysis an $n_f$-power series was used to set the effective scale without distinguishing whether the $n_f$-terms are specific to the $\{\beta_i\}$ terms; thus one cannot confirm the scheme-independence of the resultant pQCD series. However, if one improves this method, taking care that only the nonconformal $n_f$-terms associated with coupling constant renormalization are used to set the scale, one will obtain the same effective scale as that of Eq.(\ref{eq:PMCscale}).

An alternative single-scale approach~\cite{Mikhailov:2004iq, Kataev:2014jba} called the ``xBLM" approach, has been suggested based on a procedure called the  ``sequential Brodsky-Lepage-Mackenzie (seBLM)" approach ~\cite{Mikhailov:2004iq}. The goal of the seBLM approach is to improve the convergence of perturbative QCD expansions. For example, in the case of the $R$-ratio, the single scale of the xBLM approach is fixed by requiring the third-order coefficients to vanish after using the seBLM procedure; however, the fourth and higher-order coefficients remain non-zero (although small), thus leading to an unnatural perturbative series.

\section{Applications}

\subsection{Example I : $e^+e^- \to$ hadrons}

The annihilation of an electron and positron into hadrons provides one of the most precise platforms for testing the running behavior of the QCD coupling. The $R$-ratio is defined as
\begin{eqnarray}
R_{e^+ e^-}(Q)&=&\frac{\sigma\left(e^+e^-\rightarrow {\rm hadrons} \right)}{\sigma\left(e^+e^-\rightarrow \mu^+\mu^-\right)}\nonumber\\
&=& 3\sum_q e_q^2\left[1+R(Q)\right], \label{Re+e-}
\end{eqnarray}
where $Q=\sqrt s $. The pQCD approximant for $R(Q)$ up to $(n+1)$-loop level can be written as, $R_n(Q)=\sum_{i=0}^{n} {\cal C}_{i}(Q,\mu) \alpha^{i+1}(\mu)$. The expansion coefficients in the $\overline{\rm MS}$-scheme up to four-loop level can be found in Refs.\cite{Baikov:2008jh, Baikov:2010je, Baikov:2012zm, Baikov:2012zn}. In order to apply the PMC, one first transforms the calculated non-conformal $n_f$-power series into the $\{\beta_i\}$-series and applies the standard PMC multi-scale or single-scale procedures.

\begin{figure}[htb]
\includegraphics[width=0.48\textwidth]{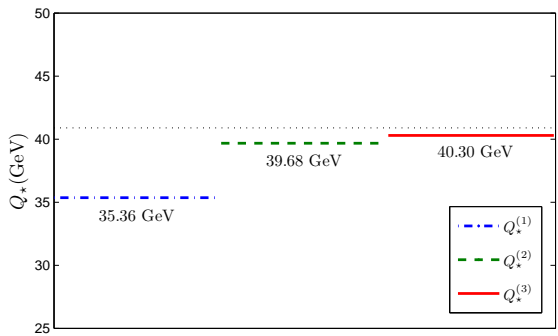}
\caption{The determined PMC scale $Q_\star$ for $R(Q)$ up to ${\rm N^{2}LL}$ accuracy. $Q^{(1)}_\star$ is at the LL accuracy, $Q^{(2)}_\star$ is at the NLL accuracy and $Q^{(3)}_\star$ is at the ${\rm N^{2}LL}$ accuracy. $Q=31.6\;{\rm GeV}$.}
\label{fig:Reescale}
\end{figure}

By using Eq.(\ref{eq:PMCscale}), the PMC scale $Q_\star$ for $R(Q)$ up to ${\rm N^{2}LL}$ precision can be determined using the four-loop pQCD prediction for $R(Q)$. The result is:
\begin{displaymath}
\ln\frac{Q_\star^2}{Q^2}=0.2249+1.6369\alpha_s(Q) +1.5559\alpha^2_s(Q).
\end{displaymath}
Numerical results are shown in Fig.(\ref{fig:Reescale}), in which $Q^{(1)}_\star$ is computed at LL, $Q^{(2)}_\star$ is for NLL and $Q^{(3)}_\star$ is for ${\rm N^{2}LL}$. For the numerical predictions, we have assumed the value for QCD mass scale $\Lambda_{\overline{\rm MS}}^{(5)} =210$ MeV, determined using the $\alpha_s$-running at four loops with $\alpha_{s,\overline{\rm MS}}(M_z)=0.1181$~\cite{Olive:2016xmw}. The results show a monotonic increase of the PMC-s single scale: $Q^{(1)}_\star < Q^{(2)}_\star < Q^{(3)}_\star$, and the difference between the two nearby values becomes smaller and smaller when more loop-terms are included. The  rapid pQCD convergence of the scale $Q_\star$ indicates that the single PMC scale converges as more loop corrections are included.

\begin{table}[htb]
\centering
\begin{tabular}[b]{cccccccc}
\hline
 & ~~$R_1$~~ & ~~$R_2$~~ & ~~$R_3$~~ & ~~$\kappa_1$~~ & ~~$\kappa_2$~~ & ~~$\kappa_3$~~ \\ \hline
~~Conv.~~ & 0.04763 & 0.04648 & 0.04617 & $7.36\%$ & $-2.43\%$ & $-0.66\%$ \\
PMC & 0.04745 & 0.04649 & 0.04619 & $6.96\%$ & $-2.03\%$ & $-0.64\%$ \\
PMC-s & 0.04745 & 0.04635 & 0.04619 & $6.96\%$ & $-2.33\%$ & $-0.34\%$ \\
\hline
\end{tabular}
\caption{Results for $R_n$ and $\kappa_n$ with various loop corrections for three scale-setting approaches. $R_0\equiv0.04437$ for all scale-settings. $Q=31.6$ GeV and $\mu=Q$.}
\label{tab:ReeRn}
\end{table}

We compare the results of $R_n(Q=31.6\;{\rm GeV})$ up to the four-loop level in Table~\ref{tab:ReeRn} using conventional scale-setting with the fixed scale $Q$ (labeled Conv.), the PMC multi-scale approach (PMC), and the PMC single-scale approach (PMC-s). We also give the results for the ratio $\kappa_{n}={(R_n-R_{n-1})}/{R_{n-1}}$ which indicates how the ``known" estimate is altered by each ``newly" available one-order-higher correction.

\begin{widetext}
\begin{center}
\begin{table}[htb]
\begin{tabular}{ c c c c c c c c}
\hline
 & ~~LO~~ & ~~NLO~~ & ~~N$^2$LO~~ & ~~N$^3$LO~~ & ~~$Total$~~ \\ \hline
~~Conv.~~ & $0.04482^{+0.00652}_{-0.00501}$ & $0.00283^{-0.00612}_{+0.00361}$ & $-0.00115^{-0.00109}_{+0.00147}$ & $-0.00033^{+0.00061}_{+0.00008}$ & $0.04617^{-0.00008}_{+0.00015}$ \\
PMC & $0.04275$ & $0.00350$ & $-0.00004$ & $-0.00002$ & $0.04619$ \\
PMC-s & $0.04292$ & $0.00339$ & $-0.00008$ & $-0.00004$ & $0.04619$ \\
\hline
\end{tabular}
\caption{The value of each loop-term (LO, NLO, N$^2$LO or N$^3$LO) for the four-loop prediction $R_3(Q)$ under three scale-setting approaches. $Q=31.6$ GeV and $\mu\in[Q/2,2Q]$. The central values for the conventional scale-setting are for $\mu=Q$. We observe that each loop term for the PMC or PMC-s is almost unchanged for $\mu\in[Q/2,2Q]$. }
\label{tab:ReeOrder}
\end{table}
\end{center}
\end{widetext}

\begin{table}[htb]
\centering
\begin{tabular}{ c c c c c c c c}
\hline
~& ~$n=1$~ & ~$n=2$~ & ~$n=3$~ \\ \hline
~$Q^{(n)}_\star$~     & ~35.36~ & ~39.68~ & ~40.30~ \\
~$Q^{(n)}_{\rm eff}$~ & ~35.36~ & ~39.02~ & ~40.29~ \\
\hline
\end{tabular}
\caption{A comparison of $Q^{(n)}_\star$ with $Q^{(n)}_{\rm eff}$ for $R_n(Q)$ up to N$^{n-1}$LO $(n=1,2,3)$ accuracy. The effective scale $Q^{(n)}_{\rm eff}$ is determined by requiring $\sum_{i=1}^{n+1} r_{i,0} \alpha^i(Q^{(n)}_{\rm eff})=R_n|_{\rm PMC}$.}
\label{tab:ReePMCscale}
\end{table}

In Table \ref{tab:ReeOrder}, we present the values of each term in the perturbative theory for the four-loop approximant $R_3(Q)$, where the errors for each loop terms by varying $\mu\in[Q/2,2Q]$ are also presented. Tables~\ref{tab:ReeRn} and \ref{tab:ReeOrder} show the PMC and PMC-s predictions are close to each other for both the total and the separate loop terms. To show this equivalency more clearly we present a comparison of $Q^{(n)}_\star$ with the effective scale $Q^{(n)}_{\rm eff}$ up to N$^{n-1}$LO $(n=1,2,3)$ accuracy in Table~\ref{tab:ReePMCscale}. Here the effective scale $Q^{(n)}_{\rm eff}$ is determined by requiring $\sum_{i=1}^{n+1} r_{i,0} \alpha^i(Q^{(n)}_{\rm eff})$ to be equal to the PMC multi-scale prediction $R_n|_{\rm PMC}$. The scales $Q^{(1)}_\star$ and $Q^{(1)}_{\rm eff}$ are by definition the same, and the high-order ones become closer to each other with increasing loop corrections; thus the PMC-s predictions are in effect, an equivalent for PMC.

\subsection{Example II: $H \to b\bar{b}$}

The pQCD prediction for decay width of the Higgs decay to a pair of bottom quarks $H\to b\bar{b}$ can be written as
\begin{eqnarray}
\Gamma(H\to b\bar{b})=\frac{3G_{F} M_{H} m_{b}^{2}(\mu)} {4\sqrt{2}\pi} [1+\tilde{R}(\mu)],
\end{eqnarray}
where $G_{F}$ is the Fermi constant, $\mu$ is the renormalization scale, and $m_{b}(\mu)$ is the $b$-quark $\overline{\rm MS}$ running mass. The pQCD prediction for $\tilde R$  takes the form $\tilde{R}_n(\mu)=\sum_{i=0}^{n} \tilde{\cal C}_i \alpha^{i+1}(\mu)$. At present, $\tilde{R}_n$ has been calculated up to four-loop level; e.g., Ref.\cite{Baikov:2005rw} gives the result, taking $\mu=M_H$, which can be run to any required perturbative scale using the RGE. We take $M_{H} =126~ {\rm GeV}$ for the Higgs mass.

\begin{figure}[h]
\includegraphics[width=0.48\textwidth]{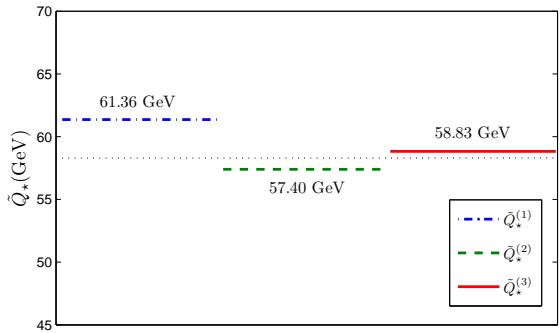}
\caption{The determined PMC scale $\tilde{Q}_\star$ for $H\to b\bar{b}$ up to ${\rm N^{2}LL}$ accuracy. $Q^{(1)}_\star$ is at the LL accuracy, $Q^{(2)}_\star$ is at the NLL accuracy and $Q^{(3)}_\star$ is at the ${\rm N^{2}LL}$ accuracy. $\mu=M_H$. }
\label{fig:Hbbscale}
\end{figure}

By using Eq.(\ref{eq:PMCscale}), the PMC-s scale $\tilde{Q}_\star$ for $\Gamma(H\to b\bar{b})$ up to ${\rm N^{2}LL}$ accuracy can be determined using four-loop prediction on $\tilde{R}(M_H)$, which reads,
\begin{displaymath}
\ln\frac{\tilde{Q}^2_\star}{M_H^2}=-1.4389-1.1847\alpha_s(M_H) +3.8753\alpha^2_s(M_H).
\end{displaymath}
As shown in Fig.(\ref{fig:Hbbscale}), the single PMC scale $\tilde{Q}_\star$ shows rapid convergence as more loop corrections are included. In contrast to $R_{e^+e^-}$, the perturbative series for $\ln{\tilde{Q}^2_\star}/{M_H^2}$ oscillates, leading to $\tilde{Q}^{(1)}_\star > \tilde{Q}^{(2)}_\star$ and $\tilde{Q}^{(2)}_\star < \tilde{Q}^{(3)}_\star$. However, similar to the case of $R_{e^+e^-}$, the absolute difference between two nearby values becomes smaller as more loop corrections are included.

\begin{table}[h]
\centering
\begin{tabular}{cccccccccc}
\hline
 & ~~$\tilde{R}_1$~~ & ~~$\tilde{R}_2$~~ & ~~$\tilde{R}_3$~~ & ~~$\tilde{\kappa}_1$~~ & ~~$\tilde{\kappa}_2$~~ & ~~$\tilde{\kappa}_3$~~ \\ \hline
~~Conv.~~ & 0.2406 & 0.2425 & 0.2411 & $18.2\%$ & $0.8\%$ & $-0.6\%$ \\
PMC & 0.2482 & 0.2404 & 0.2402       & $22.0\%$ & $-3.2\%$ & $-0.1\%$ \\
PMC-s & 0.2482 & 0.2422 & 0.2401     & $22.0\%$ & $-2.4\%$ & $-0.86\%$ \\
\hline
\end{tabular}
\caption{Results for $\tilde{R}_n$ and $\tilde{\kappa}_n$ with various loop corrections for three scale-setting approaches. $\tilde{R}_0=0.2035$ for all scale settings. $\mu=m_H$. }
\label{tab:HbbRn}
\end{table}

\begin{widetext}
\begin{center}
\begin{table}[htb]
\begin{tabular}{cccccc}
\hline
 & ~~LO~~ & ~~NLO~~ & ~~N$^2$LO~~ & ~~N$^3$LO~~ & ~~$total$~~ \\ \hline
~~Conv.~~ & $0.2031^{+0.0226}_{-0.0175}$ & $0.0374^{-0.0151}_{+0.0100}$ & $0.0019^{-0.0077}_{+0.0071}$ & $-0.0013^{-0.0005}_{+0.0020}$ & $0.2411^{-0.0007}_{+0.0016}$ \\
PMC & $0.2260$ & $0.0247$ & $-0.0093$ & $-0.0012$ & $0.2402$ \\
PMC-s & $0.2282$ & $0.0219$ & $-0.0089$ & $-0.0011$ & $0.2401$ \\
\hline
\end{tabular}
\caption{The value of each loop-term (LO, NLO, N$^2$LO or N$^3$LO) for the four-loop prediction $\tilde{R}_3(M_H)$ using  three scale-setting approaches. $\mu\in[M_H/2, 2M_H]$. The central values for the conventional scale-setting are for $\mu=M_H$. We observe that each loop term for the PMC or PMC-s is almost unchanged for $\mu\in[M_H/2, 2M_H]$. } \label{tab:HbbOrder}
\end{table}
\end{center}
\end{widetext}

We present the results for $\tilde{R}_n$ up to four loop level using various scale-setting approaches in Table~\ref{tab:HbbRn}, where the ratio $\tilde{\kappa}_n=(\tilde{R}_n -\tilde{R}_{n-1}) /{\tilde{R}_{n-1}}$. The contributions from each loop-term to the four-loop prediction $\tilde{R}_3(M_H)$ are presented in Table~\ref{tab:HbbOrder}, where the errors for each loop terms by varying $\mu\in[M_H/2, 2M_H]$ are also presented. Negative N$^2$LO or N$^3$LO values for PMC and PMC-s approaches indicate the conformal coefficients for higher-orders are negative. At four-loop level, the predictions for the decay width $\Gamma(H\to b\bar{b})$ are consistent with each other due to the excellent pQCD convergence of these scale-setting approaches.

\begin{table}[htb]
\centering
\begin{tabular}{ c c c c c c c c}
\hline
~& ~$n=1$~ & ~$n=2$~ & ~$n=3$~ \\ \hline
~$\tilde{Q}^{(n)}_\star$~     & ~61.36~ & ~57.40~ & ~58.83~ \\
~$\tilde{Q}^{(n)}_{\rm eff}$~ & ~61.36~ & ~60.04~ & ~58.72~ \\
\hline
\end{tabular}
\caption{A comparison of $\tilde{Q}^{(n)}_\star$ with $\tilde{Q}^{(n)}_{\rm eff}$ for $\tilde{R}_n$ up to N$^{n-1}$LO $(n=1,2,3)$ accuracy. The effective scale $\tilde{Q}^{(n)}_{\rm eff}$ is determined by requiring $\sum_{i=1}^{n+1} r_{i,0} \alpha^i(\tilde{Q}^{(n)}_{\rm eff})=\tilde{R}_n|_{\rm PMC}$.}
\label{tab:HbbPMCscale}
\end{table}

We present a comparison of $\tilde{Q}^{(n)}_\star$ with $\tilde{Q}^{(n)}_{\rm eff}$ for $\tilde{R}_n$ up to N$^{n-1}$LO $(n=1,2,3)$ accuracy in Table~\ref{tab:HbbPMCscale}. Here the effective scale $\tilde{Q}^{(n)}_{\rm eff}$ is determined by requiring $\sum_{i=1}^{n+1} r_{i,0} \alpha^i(\tilde{Q}^{(n)}_{\rm eff})$ to be equal to the PMC multi-scale prediction $\tilde{R}_n|_{\rm PMC}$. Similar to the previous case, the scales $\tilde{Q}^{(1)}_\star$ and $\tilde{Q}^{(1)}_{\rm eff}$ are exactly the same, and the high-order ones become more closer to each other with increasing loop corrections.

\section{Summary}

The PMC satisfies renormalization group invariance~\cite{Wu:2013ei, Wu:2014iba} and all the other self-consistency conditions required by the renormalization group~\cite{Brodsky:2012ms}. The PMC eliminates a major systematic scale uncertainty for pQCD predictions, thus greatly improving the precision of empirical tests of the Standard Model and their sensitivity to new physics. It eliminates the need to guess the renormalization scale and its range. For example, the conventional approach assigns an uncertainty of $\left(^{+1.0\%}_{-3.0\%}\right)$, $\left(^{+0.3\%}_{-1.6\%}\right)$ or $\left(^{+0.4\%}_{-0.2\%}\right)$ to the two-loop, three-loop, and the four-loop approximants of $R(Q=31.6{\rm GeV})$ by assuming the range $1/2 Q < \mu< 2 Q$, respectively; this uncertainty is greatly suppressed via using the PMC, i.e. the PMC prediction is almost unchanged for each loop term by varying $\mu\in[1/2 Q, 2 Q]$. Furthermore, as shown by Table \ref{tab:ReeOrder}, a negligible net scale error for the four-loop prediction $R_3(Q)$ under conventional scale-setting is caused by cancelations among different orders, and the scale error for each loop term is still sizable.

In its original multi-scale approach, the PMC sets the scales order-by-order; the individual scales reflect the varying virtuality of the amplitudes at each order. In this letter, to make the scale-setting procedures simpler and easier to be automatized, we have introduced a new single-scale approach (PMC-s) which achieves many of the same goals of the PMC. The PMC-s scale is a single effective scale which effectively replaces the individual PMC scales in the sense of a mean value theorem.

The PMC-s fixes the renormalization scale by directly requiring all the RG-dependent non-conformal terms up to a given order to vanish, thus it inherits most of the features of the mutli-scale approach: Its predictions are also scheme-independent due to the resulting conformal series, and the convergence of the pQCD expansion is greatly improved due to the elimination of divergent renormalon terms. As seen explicitly in Tables~\ref{tab:ReeRn}, \ref{tab:ReeOrder}, \ref{tab:HbbRn} and \ref{tab:HbbOrder}, the resulting PMC and PMC-s predictions are effectively same for both total and differential observables. Thus the PMC-s approach, with its much simpler scale-setting procedure, can be adopted as a reliable substitute for the PMC multi-scale approach, especially when one does not need detailed information at each order. \\

\noindent{\bf Acknowledgement}: We thank Yang Ma for helpful discussions. This work was supported in part by the National Natural Science Foundation of China under Grant No.11625520, and the Department of Energy Contract No.DE-AC02-76SF00515. SLAC-PUB-16917.


\begin{thebibliography}{100}

%\cite{Brodsky:2011ta}
\bibitem{Brodsky:2011ta}
  S.~J.~Brodsky and X.~G.~Wu,
  ``Scale Setting Using the Extended Renormalization Group and the Principle of Maximum Conformality: the QCD Coupling Constant at Four Loops,''
  Phys.\ Rev.\ D {\bf 85}, 034038 (2012).

%\cite{Brodsky:2011ig}
\bibitem{Brodsky:2011ig}
  S.~J.~Brodsky and L.~Di Giustino,
  ``Setting the Renormalization Scale in QCD: The Principle of Maximum Conformality,''
  Phys.\ Rev.\ D {\bf 86}, 085026 (2012).

%\cite{Brodsky:2013vpa}
\bibitem{Brodsky:2013vpa}
  S.~J.~Brodsky, M.~Mojaza and X.~G.~Wu,
  ``Systematic Scale-Setting to All Orders: The Principle of Maximum Conformality and Commensurate Scale Relations,''
  Phys.\ Rev.\ D {\bf 89}, 014027 (2014).

%\cite{Mojaza:2012mf}
\bibitem{Mojaza:2012mf}
  M.~Mojaza, S.~J.~Brodsky and X.~G.~Wu,
  ``Systematic All-Orders Method to Eliminate Renormalization-Scale and Scheme Ambiguities in Perturbative QCD,''
  Phys.\ Rev.\ Lett.\  {\bf 110}, 192001 (2013).

%\cite{Gribov:1972ri}
\bibitem{Gribov:1972ri}
 V.~N.~Gribov and L.~N.~Lipatov,
 ``Deep inelastic e p scattering in perturbation theory,''
 Sov.\ J.\ Nucl.\ Phys.\ {\bf 15}, 438 (1972).

%\cite{Altarelli:1977zs}
\bibitem{Altarelli:1977zs}
 G.~Altarelli and G.~Parisi,
 ``Asymptotic Freedom in Parton Language,''
 Nucl.\ Phys.\ B {\bf 126}, 298 (1977).

%\cite{Dokshitzer:1977sg}
\bibitem{Dokshitzer:1977sg}
 Y.~L.~Dokshitzer,
 ``Calculation of the Structure Functions for Deep Inelastic Scattering and e+ e- Annihilation by Perturbation Theory in Quantum Chromodynamics.,''
 Sov.\ Phys.\ JETP {\bf 46}, 641 (1977).

\bibitem{Wang:2014sua}
  S.~Q.~Wang, X.~G.~Wu, Z.~G.~Si and S.~J.~Brodsky,
  ``Application of the Principle of Maximum Conformality to the Top-Quark Charge Asymmetry at the LHC,''
  Phys.\ Rev.\ D {\bf 90}, 114034 (2014).

\bibitem{Wang:2016wgw}
  S.~Q.~Wang, X.~G.~Wu, S.~J.~Brodsky and M.~Mojaza,
  ``Application of the Principle of Maximum Conformality to the Hadroproduction of the Higgs Boson at the LHC,''
  Phys.\ Rev.\ D {\bf 94}, 053003 (2016).

\bibitem{GellMann:1954fq}
  M.~Gell-Mann and F.~E.~Low,
  ``Quantum electrodynamics at small distances,''
  Phys.\ Rev.\  {\bf 95}, 1300 (1954).

\bibitem{Bi:2015wea}
  H.~Y.~Bi, X.~G.~Wu, Y.~Ma, H.~H.~Ma, S.~J.~Brodsky and M.~Mojaza,
  ``Degeneracy Relations in QCD and the Equivalence of Two Systematic All-Orders Methods for Setting the Renormalization Scale,''
  Phys.\ Lett.\ B {\bf 748}, 13 (2015).

\bibitem{BjorkenPMC}
  A.~Deur, J.~M.~Shen, S.~J.~Brodsky, Guy F. de T\'eramond and X.~G.~Wu,
  ``Test of the Principle of Maximum Conformality with the strong coupling $\alpha_s$,"
  in preparation.

  %\cite{Brodsky:2002cx}
\bibitem{Brodsky:2002cx}
  S.~J.~Brodsky, D.~S.~Hwang and I.~Schmidt,
  ``Final state interactions and single spin asymmetries in semiinclusive deep inelastic scattering,''
  Phys.\ Lett.\ B {\bf 530}, 99 (2002).

  %\cite{Boer:2002ju}
\bibitem{Boer:2002ju}
  D.~Boer, S.~J.~Brodsky and D.~S.~Hwang,
  ``Initial state interactions in the unpolarized Drell-Yan process,''
  Phys.\ Rev.\ D {\bf 67}, 054003 (2003).

%\cite{Grunberg:1991ac}
\bibitem{Grunberg:1991ac}
  G.~Grunberg and A.~L.~Kataev,
  ``On Some possible extensions of the Brodsky-Lepage-MacKenzie approach beyond the next-to-leading order,''
  Phys.\ Lett.\ B {\bf 279}, 352 (1992).

%\cite{Brodsky:1995tb}
\bibitem{Brodsky:1995tb}
  S.~J.~Brodsky, G.~T.~Gabadadze, A.~L.~Kataev and H.~J.~Lu,
  ``The Generalized Crewther relation in QCD and its experimental consequences,''
  Phys.\ Lett.\ B {\bf 372}, 133 (1996).

%\cite{Brodsky:1982gc}
\bibitem{Brodsky:1982gc}
  S.~J.~Brodsky, G.~P.~Lepage and P.~B.~Mackenzie,
  ``On the Elimination of Scale Ambiguities in Perturbative Quantum Chromodynamics,''
  Phys.\ Rev.\ D {\bf 28}, 228 (1983).

%\cite{Mikhailov:2004iq}
\bibitem{Mikhailov:2004iq}
  S.~V.~Mikhailov,
  ``Generalization of BLM procedure and its scales in any order of pQCD: A Practical approach,''
  JHEP {\bf 0706}, 009 (2007).

%\bibitem{Kataev:2014jba}
\bibitem{Kataev:2014jba}
   A.~L.~Kataev and S.~V.~Mikhailov,
   ``Generalization of the Brodsky-Lepage-Mackenzie optimization within the {¦Â}-expansion and the principle of maximal conformality,''
   Phys.\ Rev.\ D {\bf 91}, 014007 (2015).

%\cite{Baikov:2008jh} {Ree4loop}
\bibitem{Baikov:2008jh}
  P.~A.~Baikov, K.~G.~Chetyrkin and J.~H.~Kuhn,
  ``Order $\alpha^4(s)$ QCD Corrections to $Z$ and tau Decays,''
  Phys.\ Rev.\ Lett.\  {\bf 101}, 012002 (2008).

%\cite{Baikov:2010je}
\bibitem{Baikov:2010je}
  P.~A.~Baikov, K.~G.~Chetyrkin and J.~H.~Kuhn,
  ``Adler Function, Bjorken Sum Rule, and the Crewther Relation to Order $\alpha_s^4$ in a General Gauge Theory,''
  Phys.\ Rev.\ Lett.\  {\bf 104}, 132004 (2010).

%\cite{Baikov:2012zn}
\bibitem{Baikov:2012zn}
  P.~A.~Baikov, K.~G.~Chetyrkin, J.~H.~Kuhn and J.~Rittinger,
  ``Adler Function, Sum Rules and Crewther Relation of Order ${\cal O}(\alpha_s^4)$: the Singlet Case,''
  Phys.\ Lett.\ B {\bf 714}, 62 (2012).

%\cite{Baikov:2012zm}
\bibitem{Baikov:2012zm}
  P.~A.~Baikov, K.~G.~Chetyrkin, J.~H.~Kuhn and J.~Rittinger,
  ``Vector Correlator in Massless QCD at Order ${\cal O}(\alpha_s^4)$ and the QED beta-function at Five Loop,"
  JHEP {\bf 1207}, 017 (2012).

%\cite{Olive:2016xmw}{pdg}
\bibitem{Olive:2016xmw}
  C.~Patrignani {\it et al.} (Particle Data Group),
  ``Review of Particle Physics,''
  Chin.\ Phys.\ C {\bf 40}, 100001 (2016).

%\cite{Baikov:2005rw}
\bibitem{Baikov:2005rw}
  P.~A.~Baikov, K.~G.~Chetyrkin and J.~H.~Kuhn,
  ``Scalar correlator at ${\cal O}(\alpha(s)^4)$, Higgs decay into $b$-quarks and bounds on the light quark masses,''
  Phys.\ Rev.\ Lett.\  {\bf 96}, 012003 (2006).

\bibitem{Wu:2013ei}
  X.~G.~Wu, S.~J.~Brodsky and M.~Mojaza,
  ``The Renormalization Scale-Setting Problem in QCD,''
  Prog.\ Part.\ Nucl.\ Phys.\  {\bf 72}, 44 (2013).

%\cite{Wu:2014iba}
\bibitem{Wu:2014iba}
  X.~G.~Wu, Y.~Ma, S.~Q.~Wang, H.~B.~Fu, H.~H.~Ma, S.~J.~Brodsky and M.~Mojaza,
  ``Renormalization Group Invariance and Optimal QCD Renormalization Scale-Setting,''
  Rep.\ Prog.\ Phys.\  {\bf 78}, 126201 (2015).

%\cite{Brodsky:2012ms}
\bibitem{Brodsky:2012ms}
  S.~J.~Brodsky and X.~G.~Wu,
  ``Self-Consistency Requirements of the Renormalization Group for Setting the Renormalization Scale,''
  Phys.\ Rev.\ D {\bf 86}, 054018 (2012).

\end{thebibliography}
\end{document}